\documentclass[runningheads,a4paper]{llncs}

\usepackage{amssymb}
\usepackage{graphicx}
\usepackage{amsmath}
\usepackage{booktabs}
\usepackage{perpage}
\usepackage{hyperref}
\MakePerPage{footnote}
\usepackage{multirow}
\usepackage{epstopdf} 
\usepackage{tikz}
\usetikzlibrary{arrows,patterns,automata,backgrounds,decorations,fit,petri,positioning,petri,shapes,calc}
\usepackage[caption=false]{subfig}
\usepackage{url}

\urldef{\mailsa}\path|n.tax@tue.nl|
\urldef{\mailsb}\path|{ilya.verenich,m.larosa}@qut.edu.au|
\urldef{\mailsc}\path|marlon.dumas@ut.ee|

\begin{document}

\mainmatter  

\title{Predictive Business Process Monitoring with LSTM Neural Networks }

\titlerunning{Predictive Business Process Monitoring with LSTM Neural Networks}

%
%
\author{Niek Tax\inst{1} \and Ilya Verenich\inst{2,3} \and Marcello La Rosa\inst{2} \and Marlon Dumas\inst{3}}

\institute{Eindhoven University of Technology, The Netherlands\\
\mailsa
\and Queensland University of Technology, Australia\\
\mailsb
\and University of Tartu, Estonia\\
\mailsc
}

%
%

\toctitle{Predictive Process Monitoring Using LSTM}
\tocauthor{Authors' Instructions}
\maketitle

\begin{abstract}
Predictive business process monitoring methods exploit logs of completed cases of a process in order to make predictions about running cases thereof.
Existing methods in this space are tailor-made for specific prediction tasks. Moreover, their relative accuracy is highly sensitive to the dataset at hand, thus requiring users to engage in trial-and-error and tuning when applying them in a specific setting. 
This paper investigates Long Short-Term Memory (LSTM) neural networks as an approach to build consistently accurate models for a wide range of predictive process monitoring tasks. 
First, we show that LSTMs outperform existing techniques to predict the next event of a running case and its timestamp. 
Next, we show how to use models for predicting the next task in order to predict the full continuation of a running case.
Finally, we apply the same approach to predict the remaining time, and show that this approach outperforms existing tailor-made methods. 


\end{abstract}

\section{Introduction}
\label{sec:introduction}

Predictive business process monitoring techniques are concerned with predicting the evolution of running cases of a business process based on models extracted from historical event logs. A range of such techniques have been proposed for a variety of prediction tasks: predicting the next activity~\cite{Becker2014}, predicting the future path (continuation) of a running case~\cite{Polato2016}, predicting the remaining cycle time~\cite{Rogge-Solti2013}, predicting deadline violations~\cite{MetzgerLISFCDP15} and predicting the fulfillment of a property upon completion~\cite{Maggi2014}.
The predictions generated by these techniques have a range of applications. For example, predicting the next activity (and its timestamp) or predicting the sequence of future activities in a case provide valuable input for planning and resource allocation. Meanwhile, predictions of the remaining execution time can be used to prioritize process instances in order to fulfill service-level objectives (e.g. to minimize deadline violations).


Existing predictive process monitoring approaches are tailor-made for specific prediction tasks and not readily generalizable. Moreover, their relative accuracy varies significantly depending on the input dataset and the point in time when the prediction is made. A technique may outperform another one for one log and a given prediction point (e.g. making prediction at the mid-point of each trace), but under-perform it for another log at the same prediction point, or for the same log at an earlier prediction point~\cite{Francescomarino15,MetzgerLISFCDP15}. In some cases, multiple techniques need to be combined~\cite{MetzgerLISFCDP15} or considerable tuning is required (e.g. using hyperparameter optimization)~\cite{Francescomarino16} in order to achieve more consistent accuracy.

Recurrent neural networks with Long Short-Term Memory (LSTM) architectures~\cite{Hochreiter1997} have been shown to deliver consistently high accuracy in several sequence modeling application domains, e.g.\ natural language processing \cite{Mikolov2013} and speech recognition \cite{Graves2013}. 
Recently, Evermann et al.~\cite{Evermann2016} applied LSTMs to predictive process monitoring, specifically to predict the next activity in a case.

Inspired by these results, this paper investigates the following questions: (i) can LSTMs be applied to a broad range of predictive process monitoring problems, and how? and (ii) do LSTMs achieve consistently high accuracy across a range of prediction tasks, event logs and prediction points?
To address these questions, the paper puts forward LSTM architectures for predicting: (i) the next activity in a running case and its timestamp; (ii) the continuation of a case up to completion; and (iii) the remaining cycle time. The outlined LSTM architectures are empirically compared against tailor-made approaches with respect to their accuracy at different prediction points, using four real-life event logs. 



The paper is structured as follows. Section~\ref{sec:related_work} discusses related work. Section~\ref{sec:preliminaries} introduces foundational concepts and notation. Section~\ref{sec:next_activity} describes a technique to predict the next activity in a case and its timestamp, and compares it against tailor-made baselines. Section~\ref{sec:full_suffix} extends the previous technique to predict the continuation of a running case. Section~\ref{sec:remaining_cycle_time} shows how this latter method can be used to predict the remaining time of a case, and compares it against tailor-made approaches. Section~\ref{sec:conclusion_future_work} concludes the paper and outlines future work directions.

\section{Related Work}
\label{sec:related_work}

This section discusses existing approaches to predictive process monitoring for three prediction tasks: time-related predictions, predictions of the outcome of a case and predictions of the continuation of a case and/or characteristics thereof.

\subsection{Prediction of time-related properties}

A range of research proposals have addressed the problem of predicting delays and deadline violations in business processes. Pika et al.~\cite{Pika2012} propose a technique for predicting deadline violations. Metzger et al.~\cite{Metzger2012,MetzgerLISFCDP15} present techniques for predicting ``late show'' events (i.e. delays between the expected and the actual time of arrival) in a freight transportation process. Senderovich et al.~\cite{Senderovich2014} apply queue mining techniques to predict delays in case executions. 

Another body of work focuses on predicting the remaining cycle time of running cases. Van Dongen et al. predict the remaining time by using non-parametric regression models based on case variables \cite{Dongen2008}. Van der Aalst et al. \cite{Aalst2011} propose a remaining time prediction method by constructing a transition system from the event log using set, bag, or sequence abstractions. Rogge-Solti \& Weske \cite{Rogge-Solti2013} use stochastic Petri nets to predict the remaining time of a process, taking into account elapsed time since the last observed event. Folino et al. \cite{Folino2012} develop an ad-hoc clustering approach to predict remaining time and overtime faults. In this paper, we show that prediction of the remaining cycle time can be approached as a special case of prediction of a process continuation. Specifically, our approach is proven to generally provide better accuracy than \cite{Aalst2011} and \cite{Dongen2008}.


\subsection{Prediction of case outcome}

The goal of approaches in this category is to predict cases that will end up in an undesirable state. 
Maggi et al. \cite{Maggi2014}, propose a framework to predict the outcome of a case (normal vs.\ deviant) based on the sequence of activities executed in a given case and the values of data attributes of the last executed activity in a case. This latter framework constructs a classifier on-the-fly (e.g.\ a decision tree or random forest) based on historical cases that are similar to the (incomplete) trace of a running case. Other approaches construct a collection of classifiers offline. For example, \cite{Leontjeva2015} construct one classifier for every possible prediction point (e.g.\ predicting the outcome after the first event, the second one and so on). Meanwhile, \cite{Francescomarino15} apply clustering techniques to group together similar prefixes of historical traces and then construct one classifier per cluster.

The above approaches require one to extract a feature vector from a prefix of an ongoing trace. De Leoni et al. \cite{Leoni2016} propose a framework that classifies possible approaches to extract such feature vectors.

In this paper, we do not address the problem of case outcome prediction, although the proposed architectures could be extended in this direction. 



\subsection{Prediction of future event(s)}
Breuker et al. \cite{Breuker2016} use probabilistic finite automaton to tackle the next-activity prediction problem, while Evermann et al.~\cite{Evermann2016} use LSTMs. Using the latter approach as a baseline, we propose an LSTM architecture that solves the next-activity prediction problem with higher accuracy than~\cite{Evermann2016} and \cite{Breuker2016}, and that can be generalized to other prediction problems.

Pravilovic et al. \cite{Pravilovic2013} propose an approach that predicts both the next activity and its attributes (e.g. the involved resource). In this paper we use LSTMs to tackle a similar problem: predicting the next activity and its timestamp. 

Lakshmanan et al. \cite{Lakshmanan2015} use Markov chains to estimate the probability of future execution of a given task in a running case. 
Meanwhile, Van der Spoel et al \cite{Spoel2012} address the more ambitious problem of predicting the entire continuation of a case using a shortest path algorithm over a causality graph. Polato et al. \cite{Polato2016} refine this approach by mining an annotated transition system from an event log and annotating its edges with transition probabilities. In this paper, we take this latter approach as a baseline and show how LSTMs can improve over it while providing higher generalizability.

\section{Background}
\label{sec:preliminaries}
In this section we introduce concepts used in later sections of this paper.

\subsection{Event logs, traces and sequences}
For a given set $A$, $A^*$ denotes the set of all sequences over $A$ and $\sigma=\langle a_1,a_2,\dots,a_n\rangle$ a sequence of length $n$; $\langle\rangle$ is the empty sequence and $\sigma_1 \cdot \sigma_2$ is the concatenation of sequences $\sigma_1$ and $\sigma_2$. $\mathit{hd}^k(\sigma)=\langle a_1, a_2, \dots, a_k\rangle$ is the prefix of length $k$ ($0 < k < n$) of sequence $\sigma$ and $tl^k(\sigma)=\langle a_{k+1},\dots,a_n\rangle$ is its suffix. For example, for a sequence $\sigma_1=\langle a,b,c,d,e\rangle$, $\mathit{hd}^2(\sigma_1)=\langle a,b\rangle$ and $\mathit{tl}^2(\sigma_1)=\langle c,d,e\rangle$.


Let $\mathcal{E}$ be the event universe, i.e., the set of all possible event identifiers, and $\mathcal{T}$ the time domain. We assume that events are characterized by various properties, e.g., an event has a timestamp, corresponds to an activity, is performed by a particular resource, etc. We do not impose a specific set of properties, however, given the focus of this paper we assume that two of these properties are the timestamp and the activity of an event, i.e., there is a function $\pi_\mathcal{T}\in \mathcal{E}\rightarrow\mathcal{T}$ that assigns timestamps to events, and a function $\pi_\mathcal{A}\in\mathcal{E}\rightarrow\mathcal{A}$ that assigns to each event an activity from a finite set of process activities $\mathcal{A}$.

An \emph{event log} is a set of events, each linked to one trace and globally unique, i.e., the same event cannot occur twice in a log. A trace in a log represents the execution of one case.

\begin{definition}[Trace, Event Log]
	A \emph{trace} is a finite non-empty sequence of events $\sigma\in\mathcal{E}^*$ such that each event appears only once and time is non-decreasing, i.e., for $1\le i < j \le |\sigma|:\sigma(i)\neq\sigma(j)$ and $\pi_\mathcal{T}(\sigma(i))\le\pi_\mathcal{T}(\sigma(j))$. $\mathcal{C}$ is the set of all possible traces. An \emph{event log} is a set of traces $L\subseteq\mathcal{C}$ such that each event appears at most once in the entire log.
\end{definition}

Given a trace and a property, we often need to compute a sequence consisting of the value of this property for each event in the trace. To this end, we lift the function $f_p$ that maps an event to the value of its property $p$, in such a way that we can apply it to sequences of events (traces).


\begin{definition}[Applying Functions to Sequences]
	\label{def:funtoseq}
	A function $f \in X \rightarrow Y$ can be lifted to sequences over $X$ using the following recursive definition: (1) $f(\langle\rangle)=\langle\rangle$;  (2) for any $\sigma\in X^*$ and $x\in X$: $f(\sigma \cdot \langle x\rangle) = f(\sigma) \cdot \langle f(x)\rangle$.

\end{definition}

Finally, $\pi_\mathcal{A}(\sigma)$ transforms a trace $\sigma$ to a sequence of its activities. For example, for trace $\sigma=\langle e_1, e_2\rangle$, with $\pi_\mathcal{A}(e_1)=a$ and $\pi_\mathcal{A}(e_2)=b$, $\pi_\mathcal{A}(\sigma)=\langle a,b\rangle$.

\subsection{Neural Networks \& Recurrent Neural Networks}
A neural network consists of one layer of \emph{inputs units}, one layer of outputs units, and multiple layers in-between which are referred to as \emph{hidden units}. The outputs of the input units form the inputs of the units of the first \emph{hidden layer} (i.e., the first layer of hidden units), and the outputs of the units of each hidden layer form the input for each subsequent hidden layer. The outputs of the last hidden layer form the input for the output layer. The output of each unit is a function over the weighted sum of its inputs. The weights of this weighted sum performed in each unit are learned through gradient-based optimization from training data that consists of example inputs and desired outputs for those example inputs. Recurrent Neural Networks (RNNs) are a special type of neural networks where the connections between neurons form a directed cycle.

\begin{figure}[t]
	\centering
	\includegraphics[width=0.57\textwidth]{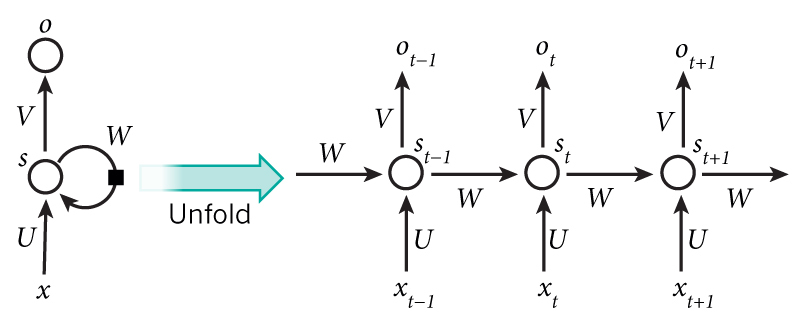}
	\caption{A simple recurrent neural network (taken from \cite{LeCun2015}).}
	\label{fig:RNNexample}
\end{figure}

RNNs can be unfolded, as shown in Figure~\ref{fig:RNNexample}. Each step in the unfolding is referred to as a time step, where $x_t$ is the input at time step $t$. RNNs can take an arbitrary length sequence as input, by providing the RNN a feature representation of one element of the sequence at each time step. $s_t$ is the hidden state at time step $t$ and contains information extracted from all time steps up to $t$. The hidden state $s$ is updated with information of the new input $x_t$ after each time step: $s_t = f(Ux_t+Ws_{t-1})$, where $U$ and $W$ are vectors of weights over the new inputs and the hidden state respectively. Function $f$, known as the activation function, is usually either the hyperbolic tangent or the logistic function, often referred to as the sigmoid function: $\mathit{sigmoid}(x)=\frac{1}{1+\mathit{exp}(-x)}$. In neural network literature the sigmoid function is often represented with the letter $\sigma$, but we will fully write $\mathit{sigmoid}$ to avoid confusion with traces. $o_t$ is the output at step $t$.\looseness=-1

\subsection{Long Short-Term Memory for Sequence Modeling}
A Long Short-Term Memory model (LSTM) \cite{Hochreiter1997} is a special Recurrent Neural Network architecture that has powerful modeling capabilities for long-term dependencies. The main distinction between a regular RNN and a LSTM is that the latter has a more complex memory cell $C_t$ replacing $s_t$. Where the value of state $s_t$ in a RNN is the result of a function over the weighted average over $s_{t-1}$ and $x_t$, the LSTM state $C_t$ is accessed, written, and cleared through controlling gates, respectively $o_t$, $i_t$, and $f_t$. Information on a new input will be accumulated to the memory cell if $i_t$ is activated. Additionally, the past memory cell status $C_{t-1}$ can be ``forgotten'' if $f_t$ is activated. The information of $C_t$ will be propagated to the output $h_t$ based on the activation of output gate $o_t$. Combined, the LSTM model can be described with the following formulas:\\
\begin{equation*}
\begin{split}
f_t=\mathit{sigmoid}(W_f\cdot[h_{t-1},x_t]+b_f)\\
i_t=\mathit{sigmoid}(W_i\cdot[h_{t-1},x_t]+b_i)\\
\tilde{C}_t=\mathit{tanh}(W_c\cdot[h_{t-1},x_t]+b_{C})\\
\end{split}
\qquad
\begin{split}
C_t=f_t*C_{t-1}+i_i*\tilde{C}_t\\
o_t=\mathit{sigmoid}(W_o[h_{t-1},x_t]+b_o)\\
h_t=o_t*\mathit{tanh}(C_t)\\
\end{split}
\end{equation*}

In these formulas all $W$ variables are weights and $b$ variables are biases and both are learned during the training phase.

\section{Next Activity and Timestamp Prediction}
\label{sec:next_activity}
In this section we present and evaluate multiple architectures for next event and timestamp prediction using LSTMs.

\subsection{Approach}
We start by predicting the next activity in a case and its timestamp, by learning an activity prediction function $f_a^1$ and a time prediction function $f_t^1$. We aim at functions $f_a^1$ and $f_t^1$ such that
$f_a^1(\mathit{hd}^k(\sigma))=\mathit{hd}^1(\mathit{tl}^k(\pi_\mathcal{A}(\sigma)))$ and $f_t^1(\mathit{hd}^k(\sigma))=\mathit{hd}^1(\mathit{tl}^k(\pi_\mathcal{T}(\sigma)))$ for any prefix length $k$.
We transform each event $e \in \mathit{hd}^k(\sigma)$ into a feature vector and use these vectors as LSTM inputs $x_1, \dots, x_k$. We build the feature vector as follows. We start with $|A|$ features that represent the type of activity of event $e$ in a so called \emph{one-hot encoding}. We take an arbitrary but consistent ordering over the set of activities $A$, and use $index\in A\rightarrow \{1,\dots,|A|\}$ to indicate the position of an activity in it. The one-hot encoding assigns the value $1$ to feature number $\mathit{index}(\pi_\mathcal{A}(e))$ and a value of $0$ to the other features. We add three time-based features to the one-hot encoding feature vector. The first time-based feature of event $e=\sigma(i)$ is the time between the previous event in the trace and the current event, i.e., $\mathit{fv}_{t1}(e)= \left\{
\begin{array}{ll}
	0  & \mbox{if } i=1, \\
	\pi_\mathcal{T}(e) - \pi_\mathcal{T}(\sigma(i-1)) & \mbox{otherwise}.
\end{array}
\right.$. This feature allows the LSTM to learn dependencies between the time differences at different points (indexes) in the process. Many activities can only be performed during office hours, therefore we add a time feature $\mathit{fv}_{t2}$ that contains the time within the day (since midnight) and $\mathit{fv}_{t3}$ that contains the time within the week (since midnight on Sunday). $\mathit{fv}_{t2}$ and $\mathit{fv}_{t3}$ are added to learn the LSTM such that if the last event observed occurred at the end of the working day or at the end of the working week, the time until the next event is expected to be longer.

At learning time, we set the target output $o_a^k$ of time step $k$ to the one-hot encoding of the activity of the event one time step later. However, it can be the case that the case ends at time $k$, in which case there is no new event to predict. Therefore we add an extra element to the output one-hot-encoding vector, which has value $1$ when the case ends after $k$. We set a second target output $o_t^k$ equal to the $\mathit{fv_{t1}}$ feature of the next time step, i.e. the target is the time difference between the next and the current event. However, knowing the timestamp of the current event, we can calculate the timestamp of the following event. We optimize the weights of the neural network with the Adam learning algorithm \cite{Kingma2015} such that the cross entropy between the ground truth one-hot encoding of the next event and the predicted one-hot encoding of the next event as well as the mean absolute error (MAE) between the ground truth time until the next event and the predicted time until the next event are minimized. 


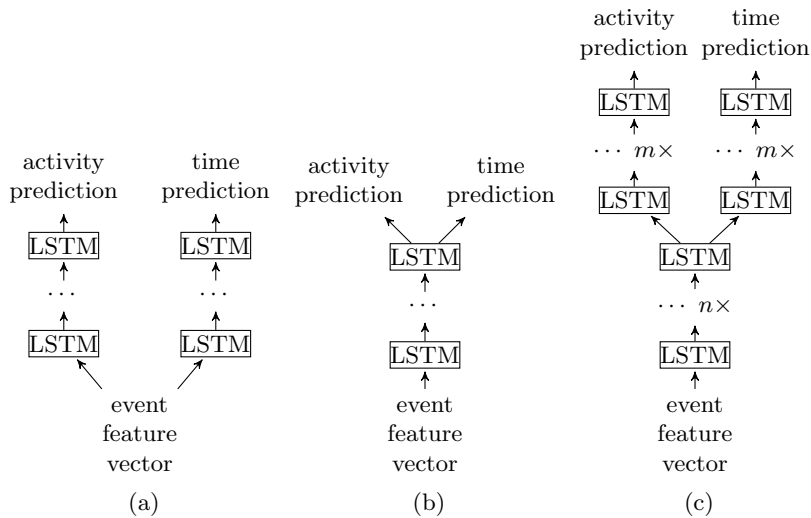
\begin{figure}[t]
	\centering
	\subfloat[]{%
		\begin{tikzpicture}
		[node distance=1.2cm,
		on grid,>=stealth',
		bend angle=20,
		auto,
		every place/.style= {minimum size=5mm},
		every transition/.style = {minimum size = 3.5mm},
		transitionH/.style={rectangle, thick, fill=black, minimum width=3mm, inner ysep=9pt }
		]
		\node (1) [align=center]{event\\ feature\\ vector};
		\node [transition] (2-1) [align=center, above left= 1.2 and 1 of 1] {LSTM}
		edge[pre] node[auto] {} (1);
		\node [transition] (2-2) [align=center, above right= 1.2 and 1 of 1] {LSTM}
		edge[pre] node[auto] {} (1);
		\node (3-1) [align=center, above = 0.65 of 2-1] {$\cdots$}
		edge[pre] node[auto] {} (2-1);
		\node (3-2) [align=center, above = 0.65 of 2-2] {$\cdots$}
		edge[pre] node[auto] {} (2-2);
		\node [transition] (4-1) [align=center, above = 0.65 of 3-1] {LSTM}
		edge[pre] node[auto] {} (3-1);
		\node [transition] (4-2) [align=center, above = 0.65 of 3-2] {LSTM}
		edge[pre] node[auto] {} (3-2);
		\node (5-1) [align=center, above= 0.9 of 4-1]{activity\\ prediction}
		edge[pre] node[auto] {} (4-1);
		\node (5-2) [align=center, above= 0.9 of 4-2]{time\\ prediction}
		edge[pre] node[auto] {} (4-2);
		\end{tikzpicture}
	}
	\subfloat[]{%
		\begin{tikzpicture}
		[node distance=1.2cm,
		on grid,>=stealth',
		bend angle=20,
		auto,
		every place/.style= {minimum size=5mm},
		every transition/.style = {minimum size = 3.5mm},
		transitionH/.style={rectangle, thick, fill=black, minimum width=3mm, inner ysep=9pt }
		]
		\node (1) [align=center]{event\\ feature\\ vector};
		\node [transition] (2) [align=center, above= 1.05 of 1] {LSTM}
		edge[pre] node[auto] {} (1);
		\node (3) [align=center, above= 0.65 of 2] {$\cdots$}
		edge[pre] node[auto] {} (2);
		\node [transition] (4) [align=center, above= 0.65 of 3] {LSTM}
		edge[pre] node[auto] {} (3);
		\node (5-1) [align=center, above left= 1 and 1 of 4]{activity\\ prediction}
		edge[pre] node[auto] {} (4);
		\node (5-2) [align=center, above right= 1 and 1 of 4]{time\\ prediction}
		edge[pre] node[auto] {} (4);
		\end{tikzpicture}
	}
	\subfloat[]{%
		\begin{tikzpicture}
		[node distance=1.2cm,
		on grid,>=stealth',
		bend angle=20,
		auto,
		every place/.style= {minimum size=5mm},
		every transition/.style = {minimum size = 3.5mm},
		transitionH/.style={rectangle, thick, fill=black, minimum width=3mm, inner ysep=9pt }
		]
		\node (1) [align=center]{event\\ feature\\ vector};
		\node [transition] (2) [align=center, above= 1.05 of 1] {LSTM}
		edge[pre] node[auto] {} (1);
		\node (3) [align=center, above = 0.65 of 2] {$\cdots$ $n\times$}
		edge[pre] node[auto] {} (2);
		\node [transition] (4) [align=center, above= 0.65 of 3] {LSTM}
		edge[pre] node[auto] {} (3);
		\node [transition] (5-1) [align=center, above left= 0.75 and 0.8 of 4] {LSTM}
		edge[pre] node[auto] {} (4);
		\node [transition] (5-2) [align=center, above right= 0.75 and 0.8 of 4] {LSTM}
		edge[pre] node[auto] {} (4);
		\node (6-1) [align=center, above = 0.65 of 5-1] {$\cdots$ $m\times$}
		edge[pre] node[auto] {} (5-1);
		\node (6-2) [align=center, above = 0.65 of 5-2] {$\cdots$ $m\times$}
		edge[pre] node[auto] {} (5-2);
		\node [transition] (7-1) [align=center, above= 0.65 of 6-1] {LSTM}
		edge[pre] node[auto] {} (6-1);
		\node [transition] (7-2) [align=center, above= 0.65 of 6-2] {LSTM}
		edge[pre] node[auto] {} (6-2);
		\node (8-1) [align=center, above= 0.9 of 7-1]{activity\\ prediction}
		edge[pre] node[auto] {} (7-1);
		\node (8-2) [align=center, above= 0.9 of 7-2]{time\\ prediction}
		edge[pre] node[auto] {} (7-2);
		\end{tikzpicture}
	}
	\caption{Neural Network architectures with single-task layers \emph{(a)}, with shared multi-tasks layer \emph{(b)}, and with $n+m$ layers of which $n$ are shared \emph{(c)}.}
	\label{fig:rnn_architectures}
	\vspace{-0.3cm}
\end{figure}

Modeling the next activity prediction function $f_a^1$ and time prediction function $f_t^1$ with LSTMs can be done using several architectures. Firstly, we can train two separate models, one for $f_a^1$ and one for $f_t^1$, both using the same input features at each time step, as represented in Figure \ref{fig:rnn_architectures} (a). Secondly, $f_a^1$ and $f_t^1$ can be learned jointly in a single LSTM model that generates two outputs, in a multi-task learning setting \cite{Caruana1997} (Figure \ref{fig:rnn_architectures} (b)). The usage of LSTMs in a multi-task learning setting has shown to improve performance on all individual tasks when jointly learning multiple natural language processing tasks, including part-of-speech tagging, named entity recognition, and sentence classification \cite{Collobert2008}. A hybrid option between the architecture of Figures \ref{fig:rnn_architectures} (a) and (b) is an architecture of a number of shared LSTM layers for both tasks, followed by a number of layers that specialize in either prediction of the next activity or prediction of the time until the next event, as shown in Figure \ref{fig:rnn_architectures} (c).

It should be noted that activity prediction function $f_a^1$ outputs the probability distribution of various possible continuations of the partial trace. For evaluation purposes, we will only use the most likely continuation.

We implemented the technique as a set of Python scripts using the recurrent neural network library \texttt{Keras} \cite{Chollet2015}. The experiments were performed on a single NVidia Tesla k80 GPU, on which the experiments took between 15 and 90 seconds per training iteration depending on the neural network architecture. The execution time to make a prediction is in the order of milliseconds. 

\subsection{Experimental setup}
In this section we describe and motivate the metrics, datasets, and baseline methods used for evaluation of the predictions of the next activities and of the timestamps of the next events. To the best of our knowledge, there is no existing technique to predict both the next activity and its timestamp. Therefore, we utilize one baseline method for activity prediction and a different one for timestamp prediction.

Well-known error metrics for regression tasks are Mean Absolute Error (MAE) and Root Mean Square Error (RMSE). Time differences between events tend to be highly varying, with values at different orders of magnitude. We evaluate the predictions using MAE, as RMSE would be very sensitive to errors on outlier data points, where the time between two events in the log is very large. 

The remaining cycle time prediction method proposed by van der Aalst et al. \cite{Aalst2011} can be naturally adjusted to predict the time until the next event. To do so we build a transition system from the event log using either set, bag, or sequence abstraction, as in \cite{Aalst2011}, but instead we annotate the transition system states with the average time until the next event. We will use this approach as a baseline to predict the timestamp of next event.

We evaluate the performance of predicting the next activity and its timestamp on two datasets. We use the chronologically ordered first 2/3 of the traces as training data, and evaluate the activity and time predictions on the remaining 1/3 of the traces. We evaluate the next activity and the timestamp prediction on all prefixes $\mathit{hd}^k(\sigma)$ of all trace $\sigma$ in the set of test traces for $2\le k<|\sigma|$. We do not make any predictions for the trace prefix of size one, since for those prefixes there is insufficient data available to base the prediction upon.

\medskip\noindent\textbf{Helpdesk dataset}
This log contains events from a ticketing management process of the help desk of an Italian software company\footnote{doi:10.17632/39bp3vv62t.1}. The process consists of 9 activities, and all cases start with the insertion of a new ticket into the ticketing management system. Each case ends when the issue is resolved and the ticket is closed. This log contains around 3,804 cases and 13,710 events.

\medskip\noindent\textbf{BPI'12 subprocess W dataset}
This event log originates from the Business Process Intelligence Challenge (BPI'12)\footnote{doi:10.4121/uuid:3926db30-f712-4394-aebc-75976070e91f} and contains data from the application procedure for financial products at a large financial institution. This process consists of three subprocesses: one that tracks the state of the application, one that tracks the states of work items associated with the application, and a third one that tracks the state of the offer. In the context of predicting the coming events and their timestamps we are not interested in events that are performed automatically. Thus, we narrow down our evaluation to the work items subprocess, which contains events that are manually executed. Further, we filter the log to retain only events of type \emph{complete}. Two existing techniques \cite{Breuker2016,Evermann2016} for the next activity prediction, described in Section \ref{sec:related_work}, have been evaluated on this event log with identical preprocessing, enabling comparison. 

\begin{table}[t]
	\scriptsize
	\centering
	\begin{tabular}{c@{\quad}c@{\quad}cccccc@{\quad}ccccc}
		\toprule
		& & & \multicolumn{5}{c}{Helpdesk} & \multicolumn{5}{c}{BPI'12 W} \\
		\multirow{2}{*}{\raisebox{-\heavyrulewidth}{Layers}} & \multirow{2}{*}{\raisebox{-\heavyrulewidth}{Shared}} & \multirow{2}{*}{\raisebox{-\heavyrulewidth}{N/l}} & \multicolumn{4}{c}{MAE in days}& \multirow{2}{*}{\raisebox{-\heavyrulewidth}{Accuracy}} & \multicolumn{4}{c}{MAE in days} &
		\multirow{2}{*}{\raisebox{-\heavyrulewidth}{Accuracy}}\\
		\cmidrule{4-7} \cmidrule{9-12} 
		&&&Prefix 2 & 4 & 6 & All & & Prefix 2 & 10 & 20 & All &  \\ 
		\midrule
		\multicolumn{13}{c}{\emph{LSTM}}\\
		4 & 4 & 100 & 3.64 & 2.79 & 2.22 & 3.82 & 0.7076 & 1.75 & 1.49& 1.02& 1.61 & 0.7466 \\
		4 & 3 & 100 & 3.63 & 2.78 & 2.21 & 3.83 & 0.7075 & 1.74 & 1.47 & 1.01 & 1.59 & 0.7479 \\
		4 & 2 & 100 & 3.59 & 2.82 & 2.27 & 3.81 & 0.7114 & 1.72 & \textbf{1.45} & 1.00 & 1.57 & 0.7497\\
		4 & 1 & 100 & 3.58 & 2.77 & 2.24 & 3.77 & 0.7074 & 1.70 & 1.46 & 1.01 & 1.59 & 0.7522 \\
		4 & 0 & 100 & 3.78 & 2.98 & 2.41 & 3.95 & 0.7072 & 1.74 & 1.47 & 1.05 & 1.61 & 0.7515\\
		3 & 3 & 100 & 3.58 & 2.69 & 2.22 & 3.77 & 0.7116 & \textbf{1.69} & 1.47 & 1.02 & 1.58 & 0.7507\\
		3 & 2 & 100 & 3.59 & 2.69 & 2.21 & 3.80 & 0.7118 & \textbf{1.69} & 1.47 & 1.01 & 1.57 & 0.7512\\
		3 & 1 & 100 & 3.55 & 2.78 & 2.38 & 3.76 & \textbf{0.7123} & 1.72 & 1.47 & 1.04 & 1.59 & 0.7525\\
		3 & 0 & 100 & 3.62 & 2.71 & 2.23 & 3.82 & 0.6924 & 1.81 & 1.51 & 1.07 & 1.66 & 0.7506\\
		2 & 2 & 100 & 3.61 & 2.64 & \textbf{2.11} & 3.81 & 0.7117 & 1.72 & 1.46 & 1.02 & 1.58 & 0.7556\\
		2 & 1 & 100 & 3.57 & \textbf{2.61} & \textbf{2.11} & 3.77 & 0.7119 & \textbf{1.69} & \textbf{1.45} & 1.01 & \textbf{1.56} & \textbf{0.7600}\\
		2 & 0 & 100 & 3.66 & 2.89 & 2.13 & 3.86 & 0.6985 & 1.74 & 1.46 & 0.99 & 1.60 & 0.7537\\
		1 & 1 & 100 & \textbf{3.54} & 2.71 & 3.16 & \textbf{3.75} & 0.7072 & 1.71 & 1.47 & \textbf{0.98} & 1.57 & 0.7486\\
		1 & 0 & 100 & 3.55 & 2.91 & 2.45 & 3.87 & 0.7110 & 1.72 & 1.46 & 1.05 & 1.59 & 0.7431  
		\vspace{0.1cm}
		\\	
		3 & 1 & 75 & 3.73 & 2.81 & 2.23 & 3.89 & 0.7118 & 1.73 & 1.49 & 1.07 & 1.62 & 0.7503\\
		3 & 1 & 150 & 3.78 & 2.92 & 2.43 & 3.97 & 0.6918 & 1.81 & 1.52 & 1.14 & 1.71 & 0.7491\\
		2 & 1 & 75 & 3.73 & 2.79 & 2.32 & 3.90 & 0.7045 & 1.72 & 1.47 & 1.03 & 1.59 & 0.7544\\
		2 & 1 & 150 & 3.62 & 2.73 & 2.23 & 3.83 & 0.6982 & 1.74 & 1.49 & 1.08 & 1.65 & 0.7511\\
		1 & 1 & 75 & 3.74 & 2.87 & 2.35 & 3.87 & 0.6925 & 1.75 & 1.50 & 1.07 & 1.64 & 0.7452\\
		1 & 1 & 150 & 3.73 & 2.79 & 2.32 & 3.92 & 0.7103 & 1.72 & 1.48 & 1.02 & 1.60 & 0.7489\\
		\multicolumn{13}{c}{\rule{0pt}{3ex} \emph{RNN}} \\
		3 & 1 & 100 & 4.21 & 3.25 & 3.13 & 4.04 & 0.6581 & & & & &\\
		2 & 1 & 100 & 4.12 & 3.23 & 3.05 & 3.98 & 0.6624 & & & & &\\
		1 & 1 & 100 & 4.14 & 3.28 & 3.12 & 4.02 & 0.6597 & & & & &\\
		\multicolumn{13}{c}{\rule{0pt}{3ex} \emph{Time prediction baselines}} \\
		\multicolumn{3}{l}{Set abstraction \cite{Aalst2011}} & 6.15 & 4.25 & 4.07 & 5.83 & - & 2.71 & 1.64 & 1.02 & 1.97 & -\\ 
		\multicolumn{3}{l}{Bag abstraction \cite{Aalst2011}} & 6.17 & 4.11 & 3.26 & 5.74 & -& 2.89&1.71 &1.07 &1.92 & -\\
		\multicolumn{3}{l}{Sequence abstraction \cite{Aalst2011}} & 6.17 & 3.53 & 2.98 & 5.67 & -& 2.89&1.69 &1.07 &1.91 &-\\
		\multicolumn{13}{c}{\rule{0pt}{3ex}  \emph{Activity prediction baselines}}\\
		\multicolumn{3}{l}{Evermann et al. \cite{Evermann2016}} & - & - & - & - & - & - & - & - & - & 0.623\\
		\multicolumn{3}{l}{Breuker et al. \cite{Breuker2016}} & - & - & - & - & - & - & - & - & - & 0.719\\
		\bottomrule
	\end{tabular}
	\caption{Experimental results for the Helpdesk and BPI'12 W logs.}
	\label{tab:helpdesk_1_lstm}
	\vspace{-0.5cm}
\end{table}

\subsection{Results}
Table \ref{tab:helpdesk_1_lstm} shows the performance of various LSTM architectures on the helpdesk and the BPI'12 W subprocess logs in terms of MAE on predicted time, and accuracy of predicting the next event. The specific prefix sizes are chosen such that they represent \emph{short}, \emph{medium}, and \emph{long} traces for each log. Thus, as the BPI'12 W log contains longer traces, the prefix sizes evaluated are higher for this log. In the table, \emph{all} reports the average performance on all prefixes, not just the three prefix sizes reported in the three preceding columns. The number of shared layers represents the number of layers that contribute to both time and activity prediction. Rows where the numbers of shared layers are $0$ correspond to the architecture of Figure \ref{fig:rnn_architectures} (a), where the prediction of time and activities is performed with separate models. When the number of shared layers is equal to the number of layers, the neural network contains no specialized layers, corresponding to the architecture of Figure \ref{fig:rnn_architectures} (b). Table  \ref{tab:helpdesk_1_lstm} also shows the results of predicting the time until the end of the next event using the adjusted method from van der Aalst et al. \cite{Aalst2011} for comparison. All LSTM architectures outperform the baseline approach on all prefixes as well as averaged over all prefixes on both datasets. Further, it can be observed that the performance gain between the best LSTM model and the best baseline model is much larger for the short prefix than for the long prefix. The best performance obtained on next activity prediction over all prefixes was a classification accuracy of 71\% on the helpdesk log. On the BPI'12 W log the best accuracy is 76\%, which is higher than the 71.9\% accuracy on this log reported by Breuker et al. \cite{Breuker2016} and the 62.3\% accuracy reported by Evermann et al. \cite{Evermann2016}. In fact, the results obtained with LSTM are consistently higher than both approaches. Even though Evermann et al. \cite{Evermann2016} also rely on LSTM in their approach, there are several differences which are likely to cause the performance gap. First of all, \cite{Evermann2016} uses a technique called \emph{embedding} \cite{Mikolov2013} to create feature descriptions of events instead of the features described above. Embeddings automatically transform each activity into a ``useful'' large dimensional continuous feature vector. This approach has shown to work really well in the field of natural language processing, where the number of distinct words that can be predicted is very large, but for process mining event logs, where the number of distinct activities in an event log is often in the order of hundreds or much less, no useful feature vector can be learned automatically. Second, \cite{Evermann2016} uses a two-layer architecture with 500 neurons per layer, and does not explore other variants. We found performance to decrease when increasing the number of neurons from 100 to 150, which makes it likely that the performance of a 500 neuron model will decrease due to overfitting. A third and last explanation for the performance difference is the use of multi-task learning, which as we showed, slightly improves prediction performance on the next activity.\looseness=-1

Even though the performance differences between our three LSTM architectures are small for both logs, we observe that most best performances (indicated in bold) of the LSTM model in terms of time prediction and next activity prediction are either obtained with the completely shared architecture of Figure \ref{fig:rnn_architectures} (b) or with the hybrid architecture of Figure \ref{fig:rnn_architectures} (c). We experimented with decreasing the number of neurons per layer to 75 and increasing it to 150 for architectures with one shared layer, but found that this results in decreasing performance in both tasks. It is likely that 75 neurons resulted in underfitting models, while 150 neurons resulted in overfitting models. We also experimented with traditional RNNs on one layer architectures, and found that they perform significantly worse than LSTMs on both time and activity prediction.

\section{Suffix Prediction}
\label{sec:full_suffix}
\enlargethispage{\baselineskip}
Using functions $f_a^1$ and $f_t^1$ repeatedly allows us to make longer-term predictions that predict further ahead than a single time step. We use $f_a^\bot$ and $f_t^\bot$ to refer to activity and time until next event prediction functions that predict the whole continuation of a running case, and aim at those functions to be such that $f_a^\bot(\mathit{hd}^k(\sigma))=\mathit{tl}^k(\pi_\mathcal{A}(\sigma))$ and $f_t^\bot(\mathit{hd}^k(\sigma))=\mathit{tl}^k(\pi_\mathcal{T}(\sigma))$
\subsection{Approach}
The suffix can be predicted by iteratively predicting the next activity and the time until the next event, until the next activity prediction function $f_a^1$ predicts the end of case, which we represent with $\bot$. More formally, we calculate the complete suffix of activities as follows:\\
$f_{a}^\bot(\sigma)= \begin{cases}
\sigma & \mbox{if } f_{a}^1(\sigma)=\bot \\
f_{a}^\bot(\sigma\cdot e), \text{with } e\in\mathcal{E},\pi_\mathcal{A}(e)=f_a^1(\sigma)\land\\\quad\pi_\mathcal{T}(e)=(f_t^1(\sigma)+\pi_\mathcal{T}(\sigma(|\sigma|))) & \mbox{otherwise}
\end{cases}$

\noindent and we calculate the suffix of times until the next events as follows:\\

\noindent$f_{t}^\bot(\sigma)= \begin{cases}
\sigma, & \mbox{if } f_{t}^1(\sigma)=\bot\\
f_{t}^\bot(\sigma\cdot e), \text{with } e\in\mathcal{E},\pi_\mathcal{A}(e)=f_a^1(\sigma)\land\\\quad\pi_\mathcal{T}(e)=(f_t^1(\sigma)+\pi_\mathcal{T}(\sigma(|\sigma|))) & \mbox{otherwise}
\end{cases}$

\subsection{Experimental Setup}
For a given trace prefix $\mathit{hd}^k(\sigma)$ we evaluate the performance of $f_a^\bot$ by calculating the distance between the predicted continuation $f_a^\bot(\mathit{hd}^k(\sigma))$ and the actual continuation $\pi_\mathcal{A}(\mathit{tl}^k(\sigma))$. Many sequence distance metrics exist, with Levenshtein distance  
being one of the most well-known ones. Levenshtein distance is defined as the minimum number of insertion, deletion, and substitution operations needed to transform one sequence into the other. 

Levenshtein distance is not suitable when the business process includes parallel branches. Indeed, when $\langle a,b \rangle$ are the next predicted events, and $\langle b, a \rangle$ are the actual next events, we consider this to be only a minor error, since it is often not relevant in which order two parallel activities are executed. However, Levenshtein distance would assign a cost of $2$ to this prediction, as transforming the predicted sequence into the ground truth sequence would require one deletion and one insertion operation. An evaluation measure that better reflects the prediction quality of is the Damerau-Levenstein distance \cite{Damerau1964}, which adds a swapping operation to the set of operations used by Levenshtein distance. Damerau-Levenshtein distance would assign a cost of $1$ to transform $\langle a,b\rangle$ into $\langle b,a\rangle$. To obtain comparable results for traces of variable length, we normalize the Damerau-Levenshtein distance by the maximum of the length of the ground truth suffix and the length of the predicted suffix and subtract the normalized Damerau-Levenshtein distance from $1$ to obtain Damerau-Levenshtein Similarity (DLS).\looseness=-1

To the best of our knowledge, the most recent method to predict an arbitrary number of events ahead is the one by Polato et al. \cite{Polato2016}. The authors first extract a transition system from the log and then learn a machine learning model for each transition system state to predict the next activity. They evaluate on predictions of a fixed number of events ahead, while we are interested in the continuation of the case until its end. We redid the experiments with their ProM plugin to obtain the performance on the predicted full case continuation.

For the LSTM experiments, we use a two-layer architecture with one shared layer and 100 neurons per layer, which showed good performance in terms of next activity prediction and predicting the time until the next event in the previous experiment (Table~\ref{tab:helpdesk_1_lstm}). In addition to the two previously introduced logs, we evaluate prediction of the suffix on an additional dataset, described below, which becomes feasible now that we have fixed the LSTM architecture. 

\medskip\noindent\textbf{Environmental permit dataset}
This is a log of an environmental permitting process at a Dutch municipality.\footnote{doi:10.4121/uuid:26aba40d-8b2d-435b-b5af-6d4bfbd7a270} Each case refers to one permit application. The log contains 937 cases and 38,944 events of 381 event types. Almost every case follows a unique path, making the suffix prediction more challenging.

\subsection{Results}
Table~\ref{tab:suffixResults} summarizes the results of suffix prediction for each log. As can be seen, the LSTM outperforms the baseline \cite{Polato2016} on all logs. Even though it improves over the baseline, the performance on the BPI'12 W log is low given that the log only contains 6 activities. After inspection we found that this log contains many sequences of two or more events in a row of the same activity, where occurrences of 8 or more identical events in a row are not uncommon. We found that LSTMs have problems dealing with this log characteristic, causing it to predict overly long sequences of the same activity, resulting in predicted suffixes that are much longer than the ground truth suffixes. Hence, we also evaluated  suffix prediction on a modified version of the BPI'12 W log where we removed repeated occurrences of the same event, keeping only the first occurrence. However, we can only notice a mild improvement over the unmodified log. 
\begin{table}[hbtp]
	\vspace*{-2mm}
		\centering
	\begin{tabular}{l|c@{\quad}c@{\quad}c@{\quad}c}
		\toprule
		Method & Helpdesk & BPI'12 W & BPI'12 W (no duplicates) & Environmental permit\\
		\midrule
		Polato \cite{Polato2016} &0.2516 & 0.0458 & 0.0336 & 0.0260\\
		LSTM & \textbf{0.7669} & \textbf{0.3533} & \textbf{0.3937} & \textbf{0.1522}\\
		\bottomrule
	\end{tabular}
	\caption{Suffix prediction results in terms of Damerau-Levenshtein Similarity.}
	\label{tab:suffixResults}
	\vspace*{-4mm}
\end{table}

\section{Remaining Cycle Time Prediction}
\label{sec:remaining_cycle_time}

\enlargethispage{0.5\baselineskip}
Time prediction function $f_t^\bot$ predicts the timestamps of all events in a running case that are still to come. Since the last predicted timestamp in a prediction generated by $f_t^\bot$ is the timestamp of the end of the case, it is easy to see that $f_t^\bot$ can be used for predicting the remaining cycle time of the running case. For a given unfinished case $\sigma$, $\hat{\sigma_t}=f_t^\bot(\sigma)$ contains the predicted timestamps of the next events, and $\hat{\sigma_t}(|\hat{\sigma_t}|)$ contains the predicted end time of $\sigma$, therefore the estimated remaining cycle time can be obtained through $\hat{\sigma_t}(|\hat{\sigma_t}|)-\pi(\sigma(|\sigma|))$.

\subsection{Experimental Setup}
We use the same architecture as for the suffix prediction experiments. We predict and evaluate the remaining time after each passed event, starting from prefix size 2. We use the remaining cycle time prediction methods of van der Aalst et al. \cite{Aalst2011} and van Dongen et al. \cite{Dongen2008} as baseline methods. 

\subsection{Results}
\begin{figure}[hbtp]
\vspace*{-2mm}
	\centering
	\includegraphics[width=0.86\textwidth]{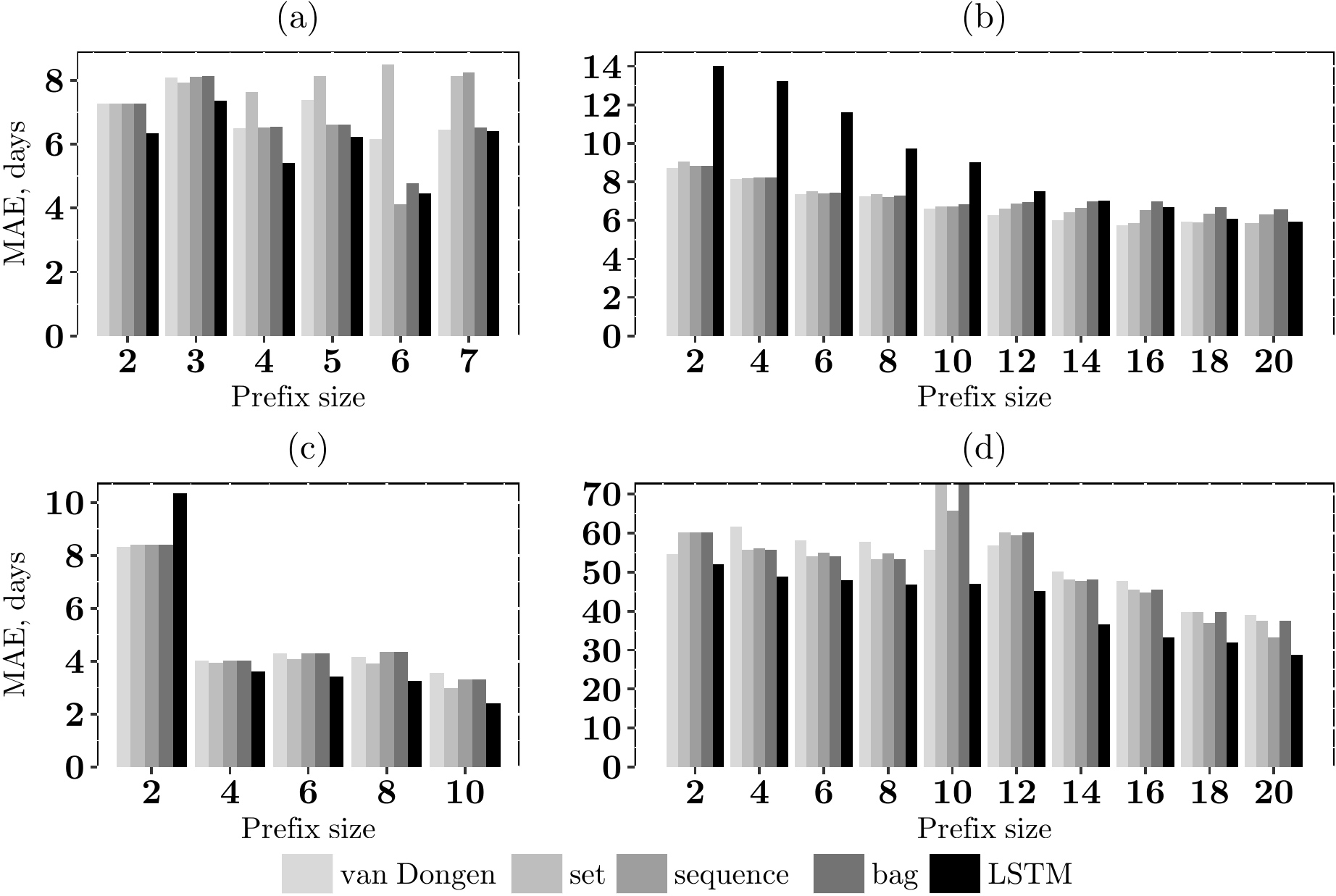}
	\caption{MAE values using prefixes of different lengths for \textit{helpdesk} (a), \textit{BPI'12 W} (b), \textit{BPI'12 W (no duplicates)} (c) and \textit{environmental permit} (d) datasets.}
	\label{fig:MaeRemainingTime}
\vspace*{-4mm}
\end{figure}

Figure~\ref{fig:MaeRemainingTime} shows the mean absolute error for each prefix size, for the four logs (Helpdesk, BPI'12 W, BPI'12 W with no duplicates and Environmental Permit). It can be seen that LSTM consistently outperforms the baselines for the Helpdesk log. An exception is the BPI'12 W log, where LSTM performs worse than the baselines on short prefixes. This is caused by the problem that LSTMs have in predicting the next event when the log has many repeated events, as described in Section \ref{sec:full_suffix}. This problem causes the LSTM to predict suffixes that are too long compared to the ground truth, and, thereby, also overestimating the remaining cycle time. We see that the LSTM does outperform the baseline on the modified version of the BPI'12 W log where we only kept the first occurrence of each repeated event in a sequence. Note that we do not remove the last event of the case, even if it is a repeated event, as that would change the ground truth remaining cycle time for the prefix.



\section{Conclusion \& Future Work}
\label{sec:conclusion_future_work}


The foremost contribution of this paper is a technique to predict the next activity of a running case and its timestamp using LSTM neural networks. We showed that this technique outperforms existing baselines on real-life data sets. Additionally, we found that predicting the next activity and its timestamp via a single model (multi-task learning) yields a higher accuracy than predicting them using separate models. We then showed that this basic technique can be generalized to address two other predictive process monitoring problems: predicting the entire continuation of a running case and predicting the remaining cycle time. We empirically showed that the generalized LSTM-based technique outperforms tailor-made approaches to these problems. We also identified a limitation of LSTM models when dealing with traces with multiple occurrences of the same activity, in which case the model predicts overly long sequences of the same event. Addressing this latter limitation is a direction for future work. 


The proposed technique can be extended to other prediction tasks, such as prediction of aggregate performance indicators and case outcomes. The latter task can be approached as a classification problem, wherein each neuron of the output layer predicts the probability of the corresponding outcome. Another avenue for future work is to extend feature vectors with additional case and event attributes (e.g.\ resources). Finally, we plan to extend the multi-task learning approach to predict other attributes of the next activity besides its timestamp.


\medskip\noindent\textbf{Reproducibility}. The source code and supplementary material required to reproduce the experiments reported in this paper can be found at \url{http://verenich.github.io/ProcessSequencePrediction}.

\medskip\noindent\textbf{Acknowledgments.} This research is funded by the Australian Research Council (grant DP150103356), the Estonian Research Council (grant IUT20-55) and the RISE{\_}BPM project (H2020 Marie Curie Program, grant 645751).


\bibliographystyle{splncs03}
\bibliography{paper}

\end{document}